\documentclass[preprint,prb,groupedaddress,showpacs]{revtex4}
\usepackage{epsfig}
\usepackage{psfrag}
\begin{document}
\title{Suppression of electron relaxation and dephasing rates in quantum dots caused by external magnetic fields}
\author{V.N. Stavrou$^{1,2}$}
\affiliation{ $^{1}$Department of Physics, State University of New York at Buffalo, New York 14260, USA\\
$^{2}$Department of Physics and Astronomy, University of Iowa, Iowa City, IA 52242, USA}
\begin{abstract}
An external magnetic field has been applied in laterally coupled  dots (QDs)
and we have studied the QD properties related to charge decoherence.
The significance of the applied magnetic field to the suppression of electron-phonon 
relaxation and dephasing rates has been explored.  
The coupled QDs have been studied by varing the magnetic field and the
interdot distance as other system parameters. Our numerical results show that the electron 
scattering rates are strongly dependent on the applied external magnetic field and the details of
the double QD configuration.
\end{abstract}
\maketitle
\date{\today}
\section{INTRODUCTION}
Relaxation and decoherence properties in electronic devices such as low-dimentional structures
and specifically in quantum dots (QDs) have attracted the interest of experimental
and theoretical studies. Fabricated coupled QDs have been suggested as candidates
for quantm bits (qubits) where the states of a single trapped electron within the coupled QDs
can perform the two states of the qubit.\cite{Barenco,Ekert,Sherwin,Tanamoto,Larionov,Hollenberg}
The investigated devices are mainly characterized by two important decoherence channels. The first
one is the coulomb interaction to the background charge fluctuation which is an extrinsic decoherence
and the second channel, an intrinsic decoherence, due to electron-phonon interaction.
The theoretical studies of the above mentioned intrinsic decoherence due to single elctron-phonon
coupling are mainly related to the single phonon emission process. Although the electron relaxation through
multi phonon processes is an important parameter at high temperature, at qubits operation temprature
($\sim$ 1T) these processes have not strong effects. \cite{Stavrou2006}
\par

It is worth mentioning that the last decade a vast of research has been published on double QDs 
dephasing \cite{Machnikowski2006}. In our previous work \cite{Stavrou2005}, we have reported the    
results of charge decoherence due to relaxation and dephasing rates caused 
by electron-phonon interaction. In the present investigation, we report the significance of
longitudinal acoustic  (LA) phonons (including the deformation and piezoelectric interactions) 
and LO phonons on relaxation and dephasing rates in laterally coupled QDs under the existence of 
magnetic fields.
The existence of an external magnetic field is of special importance for the electron 
states\cite{Jacak} due to the decrease of electron energy splitting as the magnetic field 
increases. 
The scattering rates have been found to depend
strongly on electron confinement and the interdot distance.
Acoustic phonons are only used in the calculations
of the relaxation rate due to the small electron energy splitting ($\leq$1meV). For the dephasing rates,
both the acoustic and optical phonons are considered.
\par
The paper is organized as follows. In section II, we describe the electron wavefunctions under the present 
of an external magnetic field and the models for the acoustical and optical phonon modes.
In section III, we outline the scattering theory of an electron 
which is scattered to an energetically smaller state by the emission of acoustical phonons.
The dephasing rates have been also studied by considering the emission of acoustical
and optical phonons. Section IV is devoted to the numerical results and interpretation
of suppression of decoherence due to magnetic field. Lastly, section V presents a 
summary of the present work.

%
\section{THE MODELS OF CHARGE QUBIT AND PHONONS}
\subsection{Electron wavefunctions}
We consider a single electron within two coupled  identical QDs with an interdot distance $2\alpha$.
The lateral electron motion is decoupled from the one along the QW growth. \cite{Bockel}
The Hamiltonian which describes the single electron motion which is confined in laterally 
coupled QDs is given by 
\begin {equation}
\label{hamil}
\hat{\mathcal{H}} = \hat{\mathcal{H}}_{\parallel} + \hat{\mathcal{H}_{z}}
\end {equation}
where the QW growth is denoted by the subscript ${"z"}$ and the lateral direction by ${ "\|" }$.

The external magnetic field enters the Hamiltonian via a magnetic vector potential ${\bf A}$.
By choosing the symmetric gauge ${\bf A}=\mathcal{B}\left( -y{\bf \hat e_{x}} + x{\bf \hat e_{y}}\right)/2$ 
then the magnetic 
field points to $z$ direction and it is given by ${\mathcal{\bf B}} = \nabla\times {\bf A}=\mathcal{B} {\bf \hat e_{z}}$.
The lateral  confinement is assumed to be parabolic for a single QD, therefore, the  Hamiltonian operator for the lateral directions has been considered as \cite{Jacak}
\begin {equation}
\label{hamil_xy}
\hat{\mathcal{H}}_{\parallel} = \frac{\hat{p}^{2}}{2m^{\ast}}
                              + \frac{1}{2}m^{\ast}\omega^{2}r_{\parallel}^{2}
			      - \frac{1}{2}\omega_{c}{\bf {L}}_{z} 
\end {equation}
where the operator of the $z$ component of the angular momentum is given by
\begin {equation}
\label{L_Z}
{\bf L_{z}}= -{\it i}\hbar\left[ -y\frac{\partial}{\partial x} + x\frac{\partial}{\partial y} \right]
\end {equation}
with $\hat{p}$ the quantum mechanical operator of momentum, 
$\omega_{0}$ is a parameter (in this case frequency)
describing the strength of the confinement in x-y plane,
$\omega_{c} = \mathcal{B}e/m^{*}$
and 
$\omega^{2}=\omega^{2}_{0}+(\omega_{c}/2)^{2}$.
The electron wavefunction can be separated to the
following envelope functions,
\begin {equation}
\label{envelope}
\psi(\bf{r}) = \psi_{\parallel}\left(\bf{r}_{\parallel}\right)
               \psi_{\it{z}}\left(\it{z}\right)
\end {equation}
In the case of a single QD, the one electron wavefunction can be given in terms of the principal quantum number 
$\it{n}$ $({\it{n}}={0,1,2,...})$ and the angular momentum quantum number $\it{m}$ $({\it{m}}={0,\pm1,\pm2,...})$ as
\begin {equation}
\label{psi_xy}
\psi_{\parallel}^{(n,m)}\left(\tilde{\rho}, \theta \right) = 
\sqrt{\frac{{n}!}{\pi l^{2}\left({n}+\left|{\it{m}}\right|\right)!}}
\tilde{\rho}^{\left|{\it{m}}\right|}e^{-\tilde{\rho}^{2}/2}
e^{{\it{im}\theta}}
{\mathcal{L}}_{{n}}^{\left|{\it{m}}\right|}\left(\tilde{\rho}^{2}\right)
\end {equation}
where ${\mathcal{L}}_{n}^{\left|{\it{m}}\right|}\left(\tilde{\rho}^{2}\right)$ are the Laguerre polynomials, 
and $\tilde{\rho}$ is a scaled radius, $\tilde{\rho}={\bf{r}_{\parallel}/}{\it{l}}$, with 
${\it{l}}=\sqrt{\hbar/m^{\ast}\omega_{0}}$.
The eigenvalues of the single particle are given by
\begin {equation}
\label{Vc}
E_{nm}= \left(2n+\left|{\it{m}}\right|+1\right)\hbar\omega_{0}
\end {equation} 
By using the Heaviside step function $\Theta$, the Hamiltonian, along the QW growth, takes the form 
\begin {equation}
\label{hamil_z}
\hat{\mathcal{H}_{z}} = -\frac{\hbar}{2}\partial_{z}\frac{1}{m^{\ast}(z)}\partial_{z}
                        + V_{0}\Theta\left(\left|z\right|-L_{z}\right)
\end {equation}
where $m^{\ast}(z)$ is the electron effective mass and $V_{0}$ the offset between the band 
edges well and barrier.
The wavefunction along the above mentioned direction has been considered as the
wavefunction of an infinite QW $(V_{0}\rightarrow \infty)$. Here, it has been used only the 
ground state wavefunction along the QW growth due to the strong confinement along this direction.
The ground state wavefunction is given 
by $\psi_{\it{z}}\left(\it{z}\right) = \mathcal{A} \cos\left(\pi z/2L_{z} \right)$ where $\mathcal{A}$ is a 
coefficient to be determined by normalization and $2L_{z}$ is the size of the QW. 
\par
The wavefunction of a single electron which is confined in a 2D QD and is coupled in one dimension of the x-y
plane, can be formed by superposition of two uncoupled QDs which are sited ``left'' and ``right'' of the
the origin of the frame of reference and are separated by an inter-dot distance $2\alpha$. 
The external confining potential which it is used is given by
\begin {equation}
\label{Vc}
V_{c} = \frac{1}{2}m^{\ast}\omega_{0}^{2}~min\{ \left(x-\alpha\right)^{2}+y^{2},~
                                                    \left(x+\alpha\right)^{2}+y^{2} 
                                                    \}
\end {equation}
The single electron wavefunction for the parallel plane can be given by: 
\begin {equation}
\label{superposition}
\left |\Psi_{\|}\right> = \sum_{k}{C_{k}\left |\psi_{\|,L}^{k}\right> + 
                              D_{k}\left |\psi_{\|,R}^{k}\right>}
\end {equation}
and the total wavefunction of the system of coupled QDs as described above is
\begin {equation}
\label{wavefunction}
\Psi(\bf{r}) = \Psi_{\|}\left(\bf{r_{\|}}\right)
                \psi_{\it{z}}\left(\it{z}\right)
\end {equation}
The wavefunctions in the parallel plane of the coupled QDs system are calculated 
numerically by direct diagonalization.
\subsection{Acoustic and optical phonons}
The electrons, in polar semiconductors couple to acoustical and optical phonons. 
The small electron energy splitting does not permit  any electron transition in the charge qubit via the emission of optical phonons. On the other hand, the acoustical phonons contribute to the
relaxation rates due to the small phonon energies. 
Here, we calculate the decohence rates which are caused due deformation potential and piezoelectric
acoustic phonon interaction by considering only longitudinal phonons.    
It follows the Hamiltonian which describes these interactions \cite{Mahan}: 
\begin {equation}
\label{phonons1}
H = \sum_{\bf q} \left( \frac{\hbar}{2 \rho_m V \omega_{\bf q}}
\right)^{1/2} {\mathcal{M}}({\bf q})
\rho({\bf q}) (a_{\bf q}+a_{-\bf q}^{\dagger}) \,,
\end {equation}
where $\omega_{\bf q}$ is the frequency of the phonon mode
with wavevector ${\bf q}$, $\rho_m$ is the mass density of the
host material, $V$ is the volume of the sample, $a_{\bf q}$ and $a_{-\bf
q}^\dagger$ are phonon annihilation and creation operators, and $\rho({\bf
q})$ is the electron density operator. The term $\mathcal{M}({\bf q})$ is given by
\begin {equation}
\label{phonons2}
\mathcal{M}({\bf q}) = D\left| \bf{q} \right| + {\it i}\mathcal{M}_{\lambda}(\hat {q})
\end {equation}
The first term of the above equation represents the deformation potential
interaction with deformation potential D and the second part,
which is imaginary, is the piezoelectric interaction. \cite{MahanPAP1}
For zincblende crystals (e.g. GaAs), the term $\mathcal{M}_{\lambda}(\hat {q})$ 
can get the form \cite{Bruus1993}
\begin {equation}
\label{PZ}
{\mathcal{M}}^{pz}_{\lambda} (\hat{\bf q}) = 2 e \ e_{14}\left( \hat{q}_{x}
\hat{q}_{y} \xi_{z} + \hat{q}_{y} \hat{q}_{z} \xi_{x} + \hat{q}_{x}
\hat{q}_{z} \xi_{y} \right)
\end {equation}
where $e$ is the electronic charge, $e_{14}$ is the piezoelectric constant,
and $\xi$ is the unit polarization vector.
\par
The fact that the energy difference between the electron states in coupled quantum dots 
is quite small (a few meV)
does not allow optical phonon transitions due to the conservation of energy
(optical phonon energy is $\sim$ 36 meV in GaAs).
The optical phonons play a role in dephasing rates as we demonstrate in the next section.
By using the bulk phonon approximation and neglecting the interface phonon modes 
\cite{Essex,Ridley96},
the electron-phonon interaction due to LO phonons is thus given by \cite{Mahan}
\begin{equation}
H_{OP} = \sum_{\bf q} \frac{M}{q\sqrt{V}}
\rho({\bf q}) (a_{\bf q}+a_{-\bf q}^{\dagger})
\end{equation}
and
\begin {equation}
\label{Coupling2}
M^{2} = 2\pi e^{2}\hbar\omega_{LO}
        \left( \frac{1}{\epsilon_\infty{}} -
\frac{1}{\epsilon_{s}} \right)    
\end {equation}
where $\omega_{LO}$ is the longitudinal optical frequency,
$\epsilon_{s}$ and $\epsilon_{\infty}$ are the static and high frequency
dielectric constant of the host material.
\par
Having described the electronic states in coupled QDs and the relevant types of phonon, 
we are now ready to calculate the relaxation and dephasing rates.
The next section is devoted to Fermi's golden rule and dephasing rates.   
%
%
\section{THEORY OF RELAXATION AND DEPHASING RATES}
%
%
The relaxation rate between an initial $\left| \Psi^{(I)}\right>$ and a final state $\left| \Psi^{(F)}\right>$ 
associated with phonon emission (+) (or absorption (-)) is determined by Fermi's golden rule:
\begin {eqnarray}
\label{fermi}
\Gamma  &=&  \frac{2\pi}{\hbar}
\sum_{\bf{q}} 
\left| \left< \Psi^{(F)}({\bf{r}})\left|H^{\it{int}}
\right| \Psi^{(I)}({\bf{r}}) \right> \right|^{2}  
\delta\left(E_{F}-E_{I} \pm E_{\bf{q}} \right)
\nonumber\\ & &
\left(N_{B}(E_{\bf{q}},T_{lat})+\frac{1}{2} \pm \frac{1}{2}   \right)
\end {eqnarray}
where the labels `I' and 'F' denote the initial and final electron states 
respectively. $N_{B}$ is the Bose-Einstein distribution function for phonons with 
lattice temperature $T_{lat}$. It is worth mentioning that in our calculations
we assumed $T_{lat}=0$ and the phonon absorption can be neglected.
\par
It is also worth mentioning that relaxation is not the only way charge qubits can be decohered.
If the energy difference between the two charge states fluctuates, phase information will
get lost and decoherence occurs.  The density
operator of an electron in a boson bath is given in \cite{Palma96,Duan}
\begin{eqnarray} 
\label{density_matrix}
\rho(t) = \left( \begin{array}{cc} \rho_{00}(0) & \rho_{01}(0)e^{-B^{2}(\Delta
t)+i\varepsilon \Delta t/\hbar} \\ 
\rho_{10}(0)e^{-B^{2}(\Delta t)-i\varepsilon \Delta t/\hbar} & \rho_{11}(0)
\end{array} \right)
\end{eqnarray} 
where $\varepsilon$ is the energy splitting between the electron energy
levels.  In short, pure dephasing cause a decay in the off-diagonal element of
the density matrix for the two-level system that makes up the charge qubit
\cite{Palma96,Duan}:
\begin {equation}
\rho_{01}(t)\sim \rho_{01}(0)e^{-B^{2}(t)} \,,
\label{eq:rho_10}
\end {equation}
where the exponent function $B^{2}(t)$ is defined by
\begin {equation}
\label{spectral}
B^{2}(t) = \frac{V}{\hbar^{2} \pi^{3}} \int{d^{3}{\bf q} \frac{|g({\bf
q})|^{2}}{\omega_{\bf q}^2} \sin^{2} \frac{\omega_{\bf q} t}{2} \coth
\frac{\hbar \omega_{\bf q}}{2k_{B} T}} \,.
\end {equation}
For acoustic phonons, we choose frequencies $\omega_{\bf q} = qc_{s}$ for the relevant
branches, while for longitudinal optical phonons, we choose $\omega_{\bf q} =
\omega_{LO}$.  The coupling constants $g({\bf q})$ due to deformation potential, piezoelectric and optical phonons are respectively given by 
\begin {eqnarray}
\label{dephas_Elem_D}
g_{\rm def} ({\bf q}) = D \sqrt{\frac{\hbar q}{2\rho c_{s}V}}
\mathcal{I}({\bf q}) \,, \\
\label{dephas_Elem_P}
g_{\rm piezo} ({\bf q}) = {\mathcal{M}}^{pz}_{\lambda}({\bf q}) \ 
\sqrt{\frac{\hbar}{2\rho c_{s}V}}
\mathcal{I}({\bf q}) \,, \\
\label{dephas_Elem_LO}
g_{\rm polar}({\bf q}) = \frac{M}{q\sqrt{V}}
                             \mathcal{I}({\bf q}) \,,
\end {eqnarray}
where $\mathcal{I}({\bf q})$ is given by
\begin {eqnarray}
\label{I_q}
\mathcal{I}({\bf q}) = \frac{1}{2}\left(\left<
\Psi^{-}({\bf{r}})\left|e^{\mp{\it{i}}{\bf{q \cdot r}}}
\right| \Psi^{-}({\bf{r}})\right>
- \left< \Psi^{+}({\bf{r}})\left|e^{\mp{\it{i}}{\bf{q \cdot r}}}
\right| \Psi^{+}({\bf{r}})\right>
\right)
\end {eqnarray}
the symbols $(\pm)$ refer to the two states for the double dot charge qubit.  The matrix integrals in 
this study are carried out using the Monte-Carlo algorithms.
%
%
\section{RESULTS AND DISCUSSIONS}
%
We firstly calculate the relaxation rates as a function
of an external magnetic field $\mathcal{B}$, 
for an electron which scatters from 
the first excited to ground state associated by the 
emission of acoustical phonons. 
In all our calculations, the quantum well width is fixed 
to the value of $2L_{z}=10~nm$ and the confinement lengths in the x an y directions are
$0.5~\mu m$.
\par
\begin{figure}[]
\begin{center}
\psfrag{BB}{$\mathcal{B}$(Tesla)}
\includegraphics[trim= 0 0 0 -45, width=3.0in]{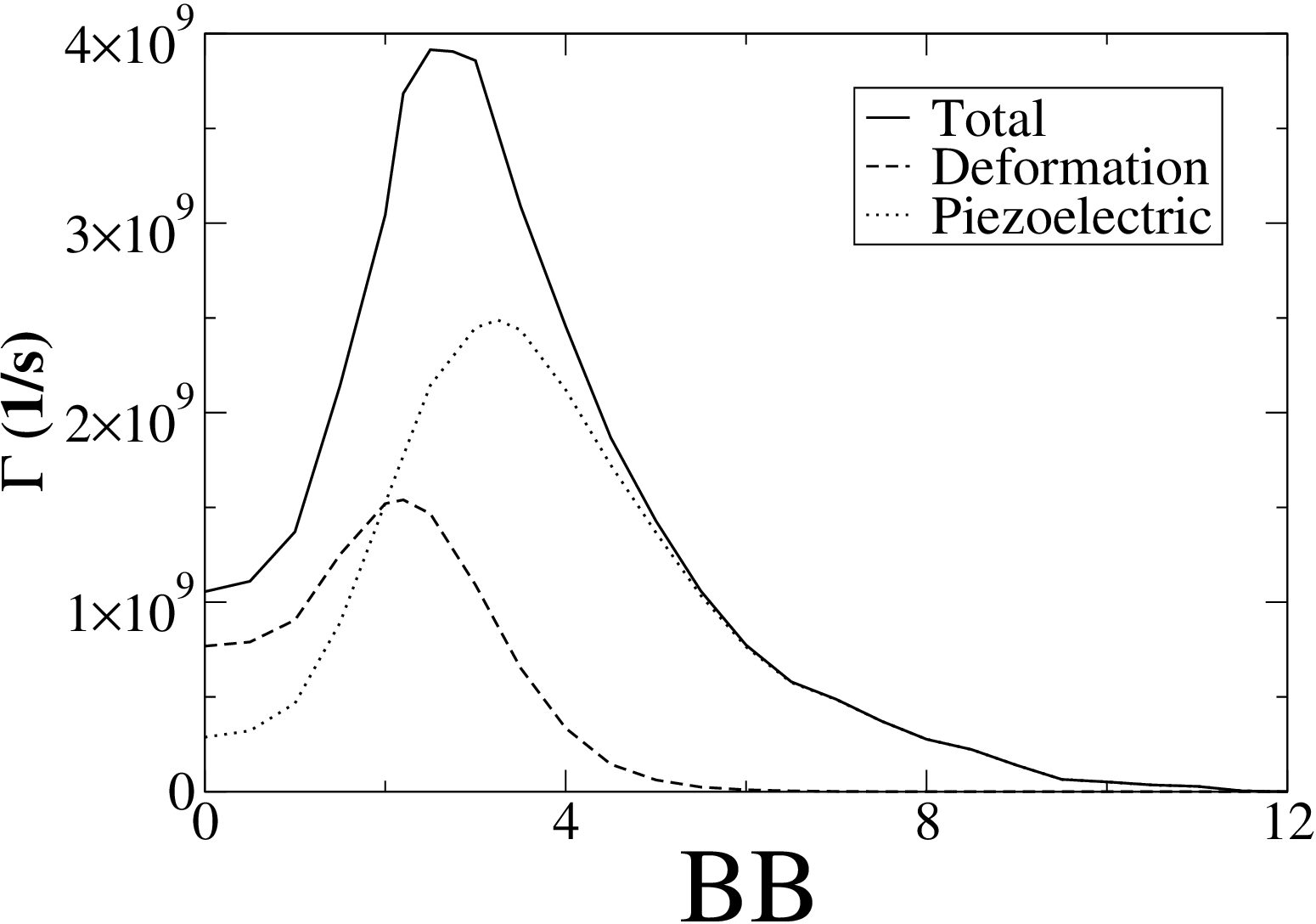}
\protect\caption{{Relaxation rates of an electron, which scatters from the first excited state to ground state, versus the external magnetic field. The confinement strength is $\hbar\omega_{0}= 3~meV$, half the interdot distance $\alpha=20~nm$ and QW width $2L_{z}=10~nm$.}.}
\label{fig_1}
\end{center}
\end{figure}

As it appears in Fig. \ref{fig_1}, for small magnetic field 
the relaxation rates due to deformation 
interaction is larger than the one due to piezoelectric interaction.
As the field $\mathcal{B}$ increases the piezoelectric coupling 
becomes the dominate contributor due to the different wavevector 
dependence in the deformation and piezoelectric matrix elements.
For the deformation potential, the dependence of matrix elements 
on wavevector is related to $\sqrt{q}$ while for the piezoelectric 
coupling is related to $1/\sqrt{q}$. 
The relaxation rates for small magnetic field increase up to 
a maximum value and afterwards
decrease. This resonance reflects the existence of large electron
wavefunctions of the charge qubit (for $\mathcal{B} \approx$3~Tesla). 
\par
\begin{figure}[]
\begin{center}
\includegraphics[trim= 0 0 0 -45, width=3.0in]{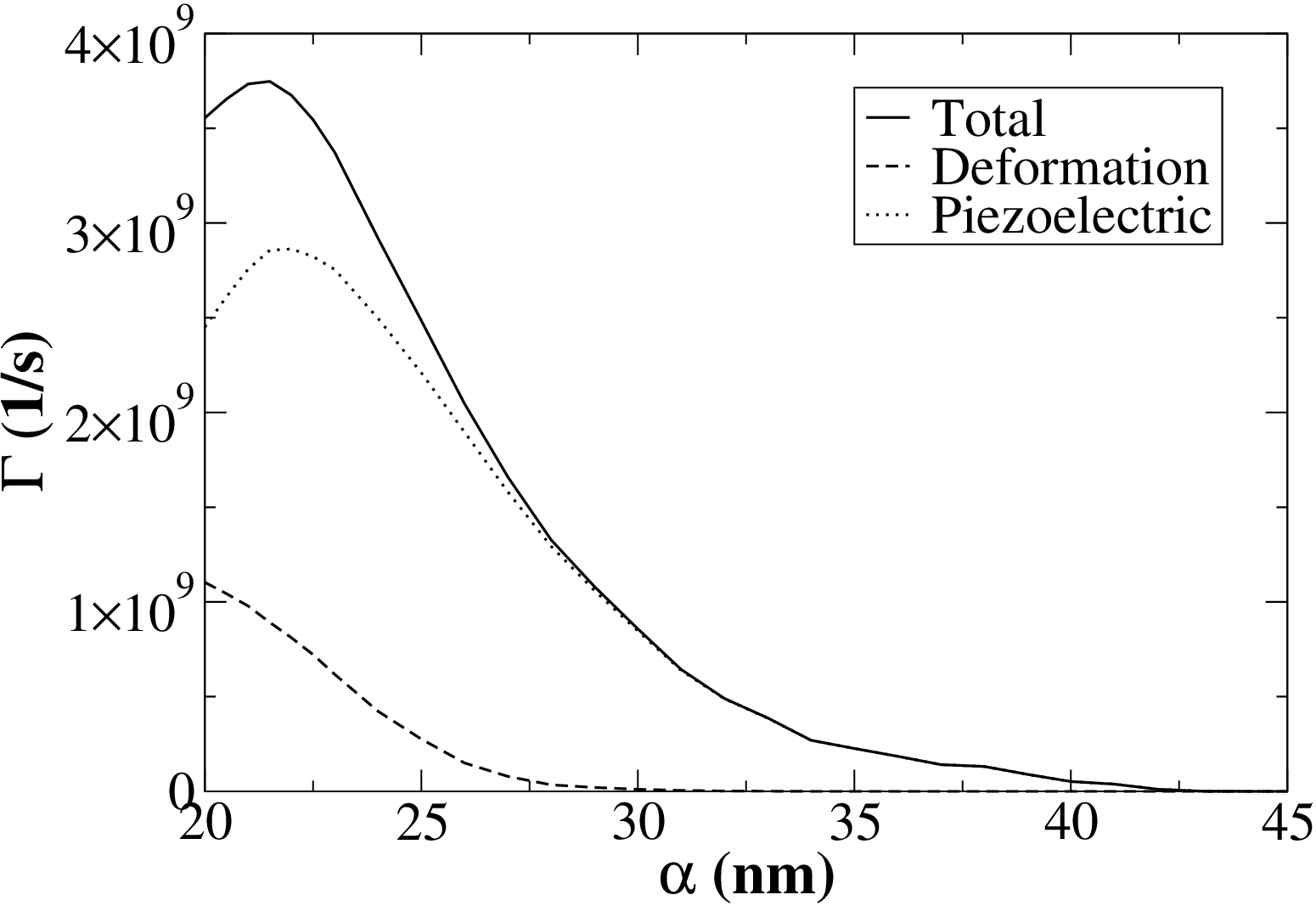}
\protect\caption{Relaxation rates of an electron versus half the interdot distance  $\alpha$. 
The confinement strength is $\hbar\omega_{0} = 3~meV$, the magnetic field $\mathcal{B}= 3$ Tesla and QW width $2L_{z}=10~nm$.
}
\label{fig_2}
\end{center}
\end{figure}

Fig. \ref{fig_2} presents the electron relaxation rates as a function the half the 
interdot distance for a fixed $\mathcal{B}=3$ Tesla and 
$\hbar\omega = 3~meV$.
As the interdot distance increases the rates decrease due to the small 
energy splitting which results small phonon density of states. Thus 
the relaxation rates become small as the interdot distance increases.
For $\alpha$ close to $21.5~nm$, it is obvious that the rates get 
a maximum value due to large wavefunctions. The different 
dependence of matrix elements on the phonon wavefuction for
the deformation potential and piezoelectric interaction results
the different behavior of the relaxation rates for the above
mentioned interactions.  
\par
\begin{figure}[]
\begin{center}
\includegraphics[trim= 0 0 0 -45, width=3.0in]{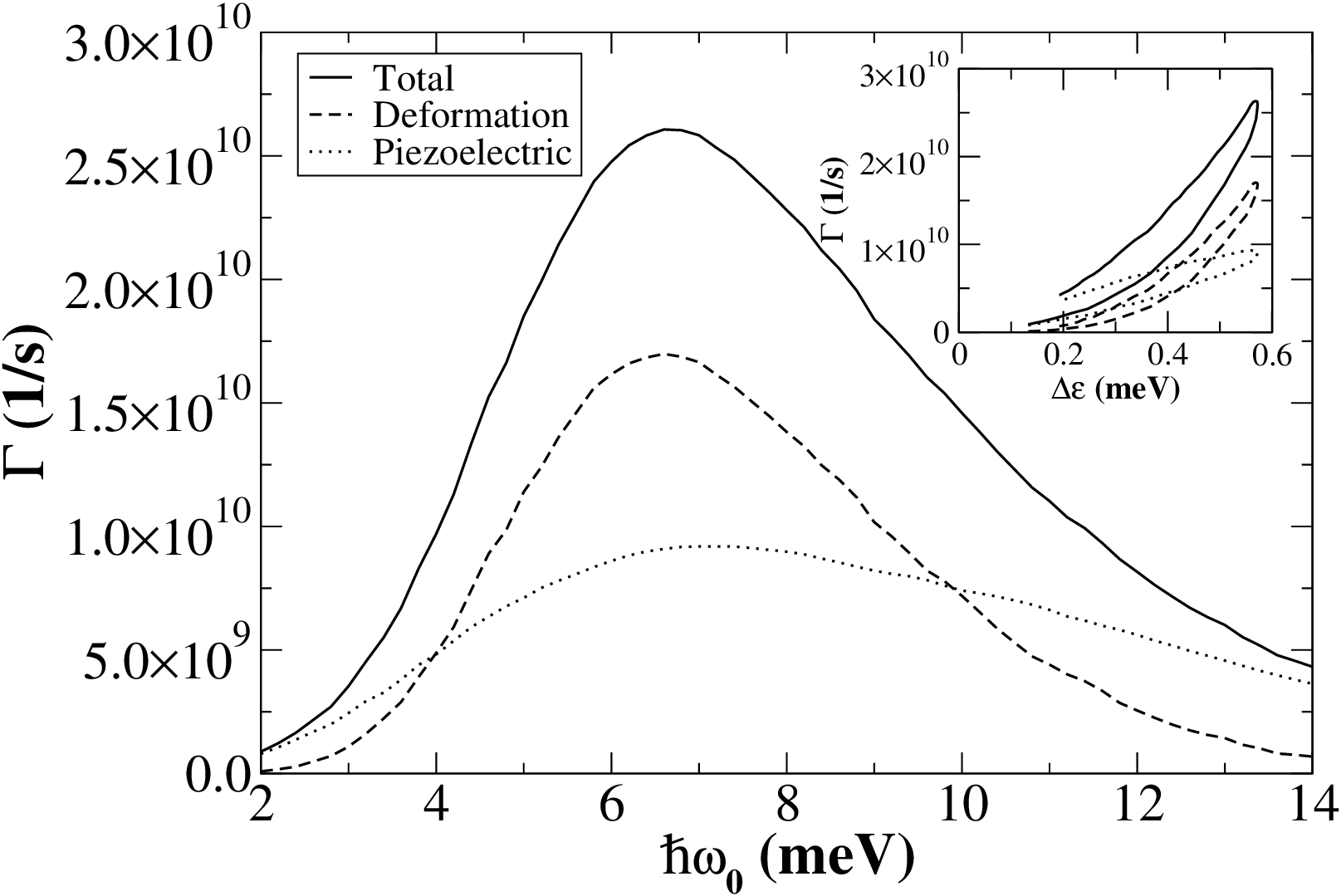}
\protect\caption{Relaxation rates of an electron versus the strength of the confinement $\hbar\omega_{0}$. Electron relaxation rates as a function of the energy splitting between the first excited state and the ground state is given in the inset. The half the interdot distance is $\alpha=20~nm$, magnetic field $\mathcal{B}=3$ Tesla and QW width $2L_{z}=10~nm$.}
\label{fig_3}
\end{center}
\end{figure}

The dependence of the relaxation rates on the electron confinement
strength is shown in  Fig. \ref{fig_3}.
The rates increase as the energy splitting between the first excited
and ground state increases. When the electron strength of the 
electron confinement reaches the value $\hbar\omega=6.5$ meV, the
relaxation rates get a maximum value. Increasing the electron 
confinement strength, the energy splitting becomes small
and the relaxation rates decrease as a consequence of the 
energy splitting degradation.
The dependence of relaxation rates on the energy splitting 
is shown in inset \ref{fig_3}. Which reflects the fact that for a given energy splitting,
correspond two values of
the strength of the confinement. \cite{Stavrou2005} 
As a result there are two values of relaxation rates for 
each energy splitting.
The two different types of electron-phonon interactions
produce different contribution to the total relaxation rates
and can be interpreted in the same manner as in Fig. \ref{fig_1}. 
\par
\begin{figure}[]
\begin{center}
\includegraphics[trim= 0 0 0 -45, width=3.0in]{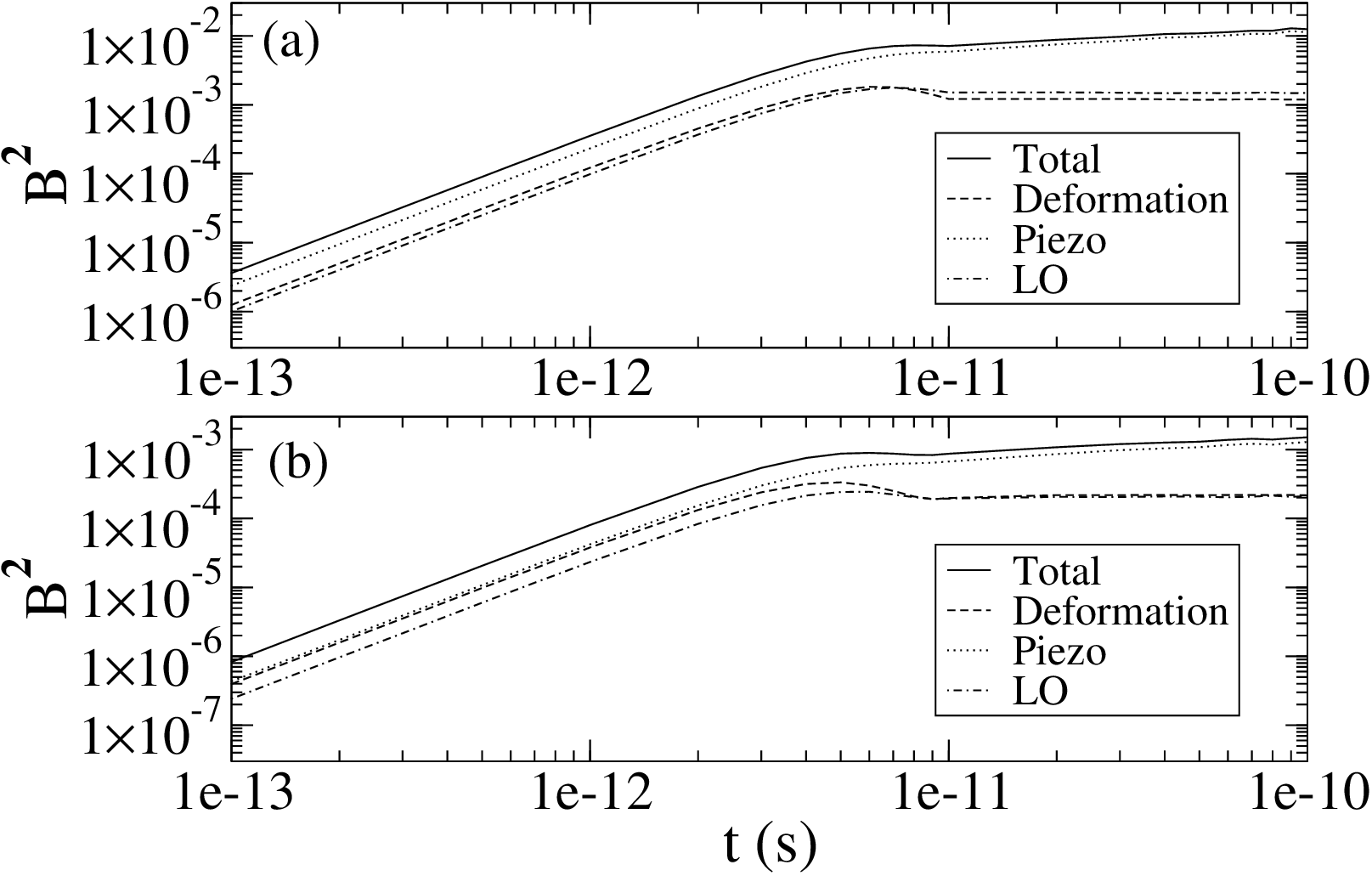}
\protect\caption{Dephasing factor $B^2(t)$ as a function of time $t$. The dephasing factor due to deformation potential, piezoelectric interaction, LO and the total rates are represented by dashed, dotted, dashed-dotted and straight line respectively.  The strength of the confinement is $\hbar\omega_{0}=3$ meV, half the interdot distance is $\alpha = 20$ nm, with a)  $\mathcal{B}=0$ Tesla and b) $\mathcal{B}=3$ Tesla.}
\label{fig_4}
\end{center}
\end{figure}

The second part of our investigation of decoherence in charge qubits is the 
evaluation of dephasing factor and its dependence on an external 
magnetic field. 
Here, we calculate the dephasing effects from both acoustic (deformation, piezoelectric) and optical phonons.
Fig. \ref{fig_4} shows the temporal dependence for two different values of an external magnetic field.
The curves in Fig. \ref{fig_4} rapidly increase for the first 10 ps and and for later times they saturate. As a result $B^{2}(t)$ depends only very slowly on time after 100~ps. The interaction between the qubit electron and the acoustic phonon bath causes the fast increasing of dephasing in a period of time less than 100~ps. Mathematically, the very fast time dependence of dephasing is due to the trigonometric 
dependence on phonon frequencies and time [see Ref.~\onlinecite{Stavrou2005}]. For larger time is practically flat and can be considered constant after 100~ps. A
constant dephasing factor will not produce a decaying signal in terms of, for
example, oscillations in electrons.  Instead, it simply reduces the contrast
in the charge oscillation.  This can be seen easily from
Eq.~\ref{density_matrix}.  The presence of a constant $\exp(-B^2)
\sim \exp(-0.05)$ simply reduces the magnitude of $\rho_{01}$ by a constant
factor of 0.05, which is not a particularly large suppression (though
significant in terms of fault tolerant quantum computing).
   
\begin{figure}[]
\begin{center}
\psfrag{BB}{$\mathcal{B}$(Tesla)}
\includegraphics[trim= 0 0 0 -45, width=3.0in]{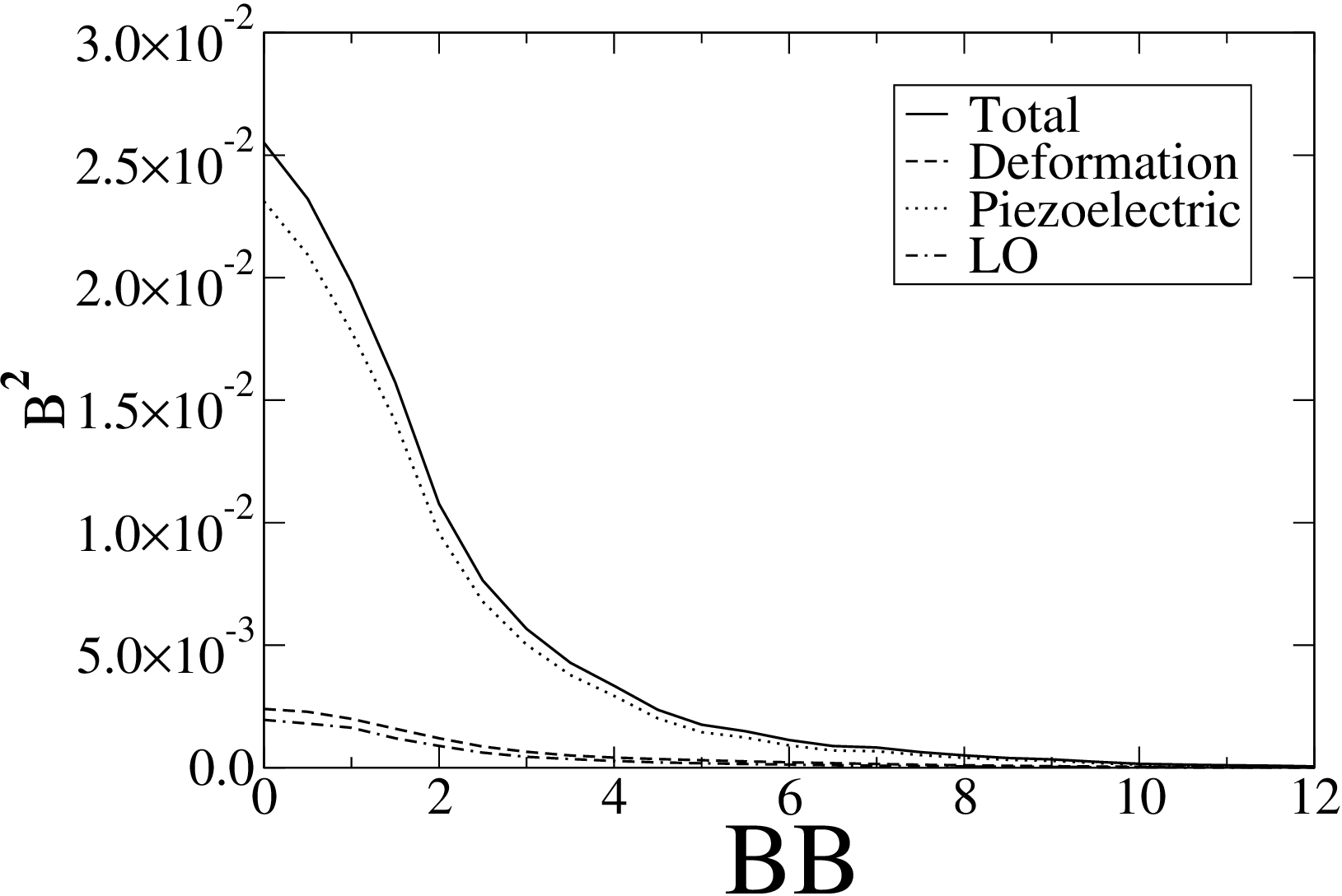}
\protect\caption{Dephasing factor $B^2(t)$ as a function of the magnetic field.
The time $t$ is chosen to be 60 ps. The solid line represents dephasing rates due electron-acoustic phonon interactions. The dephasing rates due to deformation potential, piezoelectric interaction and polar interaction with optical phonons  are represented by dashed, dotted, and dash-dotted line respectively. The strength of the confinement is $\hbar\omega_{0}=3$ meV, half the interdot distance $\alpha=20~nm$ and QW width $2L_{z}=10~nm$.}
\label{fig_5}
\end{center}
\end{figure}

The temporal behavior of the dephasing factor has an interesting feature for zero and a none zero magnetic field. 
$B^{2}(t)$ decreases as
$\mathcal{B}$ increases (Fig. \ref{fig_4}a,b) due to  the dependence of the matrix elements on the initial and final wavefunctions under the presence of an external magnetic field.
Fig. \ref{fig_5} shows the dephasing rates as a function 
of an external magnetic field. As the field $\mathcal{B}$ increases the quantity $\mathcal{I}({\bf q})$ in
Eq.(\ref{I_q}) decreases as a consequence of the  smaller differences in the matrix elements
involved in Eq. (\ref{I_q}). For large magnetic fields the rates go to zero.
\par
Finally, we calculate the dephasing rates as a function of the half the 
interdot distance for a fixed external magnetic field and electron 
confinement strength. As in the case of the relaxation rates, the
dephasing rates decrease by increasing the interdot distance due
to the fact that when the two QDS are well separated then the 
integral difference in Eq.(\ref{I_q}) goes down quickly. Consequently, the dephasing 
factor $B^{2}(t)$ undergoes suppression by increasing the separation distance.
%
%
%
\section{CONCLUSIONS}
%
We have studied the phonon-induced single electron relaxation and
 dephasing rates in laterally coupled QDs with the presence of an 
external magnetic field. 
The relaxation and dephasing rates have been calculated 
for different system parameters such as interdot distance, 
strength of electron confinement and magnetic field.
In the case of zero magnetic field \cite{Stavrou2005},
the rates could be enhanced for some double dot configurations. 
This enhancement of the rates could be easily suppressed by the existence of an external magnetic field.
Our results show that the magnetic field is of crucial 
importance in the study the decoherence in charge qubits due to the suppression
of electron relaxation and dephasing rates. 
%
\section{ACKNOWLEDGMENT}
%
The author would like to thank Prof. X. Hu for fruitful discussions on quantum computing.
The work is supported in part by NSA and ARDA under ARO contract
No.~DAAD19-03-1-0128.


%
%
\newpage

\newpage

\newpage
\begin{figure}[]
\begin{center}
\includegraphics[trim= 0 0 0 -45, width=3.0in]{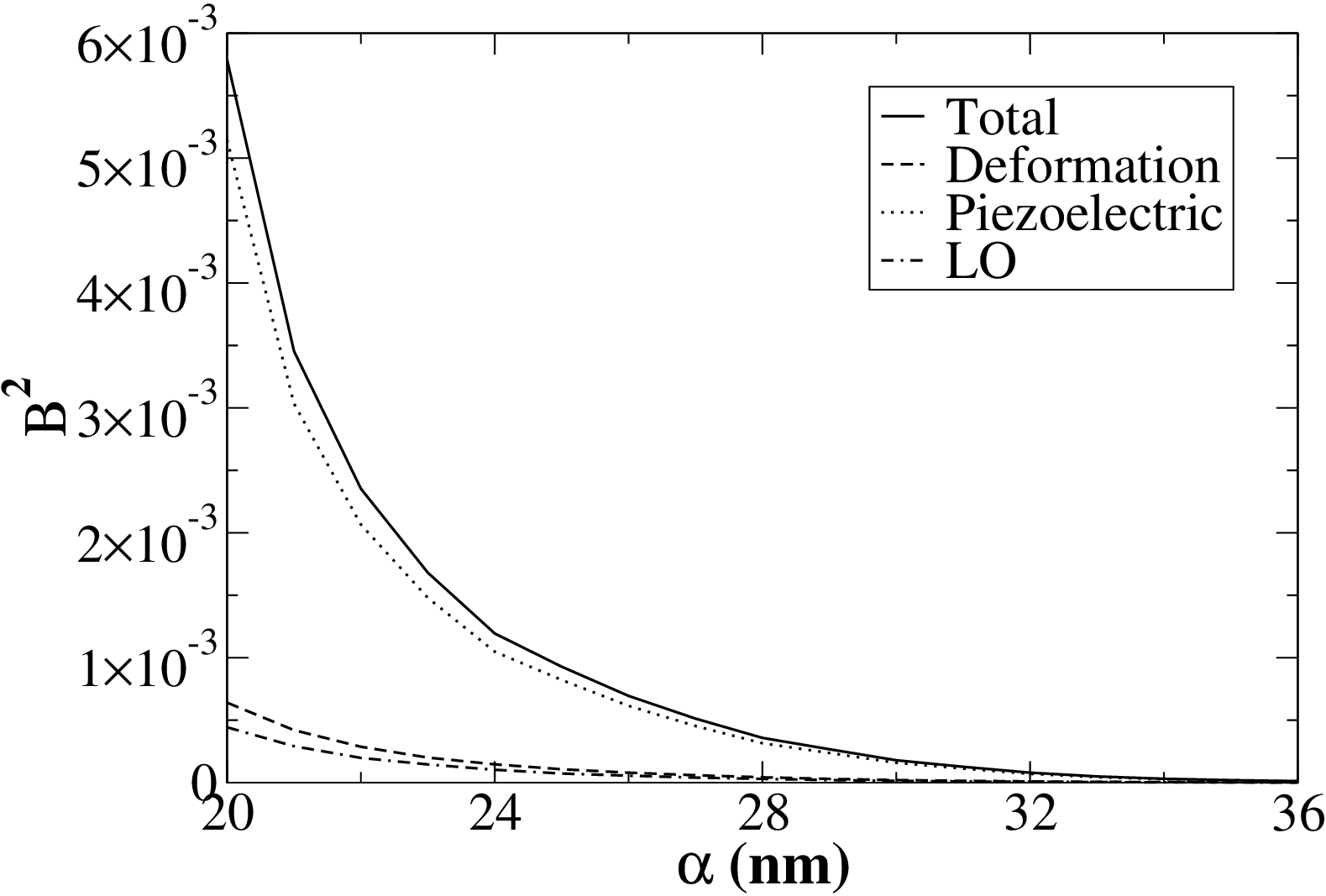}
\protect\caption{Dephasing factor $B^2(t)$ as a function of the half interdot distance $\alpha$.  The time $t$ is chosen to be 60 ps.  The solid line represents dephasing rates due electron-acoustic phonon interactions.The dephasing rates due to deformation potential, piezoelectric interaction, and polar interaction with optical phonons are represented by dashed, dotted, and dash-dotted line respectively. The strength of the confinement is $\hbar\omega_{0}=3$ meV, $\mathcal{B}= 3$ Tesla and QW width $2L_{z}=10~nm$.}
\label{fig_6}
\end{center}
\end{figure}

\end{document}